# High Precision Magnetic Linear Dichroism Measurements in (Ga,Mn)As

N. Tesařová[a)], J. Šubrt, P. Malý and P. Němec

*Faculty of Mathematics and Physics, Charles University in Prague, Ke Karlovu 3, 121 16 Prague 2, Czech Republic*

C. T. Ellis, A. Mukherjee and J. Cerne

*Department of Physics, University at Buffalo, The State University of New York, Buffalo, New York 14260, USA*

Investigation of magnetic materials using the first-order magneto-optical Kerr effects (MOKE) is well established and is frequently used in the literature. On the other hand, the utilization of the second-order (or quadratic) magneto-optical (MO) effects for the material research is rather rare. This is due to the small magnitude of quadratic MO signals and the fact that the signals are even in magnetization (i.e., they do not change a sign when the magnetization orientation is flipped), which makes it difficult to separate second-order MO signals from various experimental artifacts. In 2005 a giant quadratic MO effect - magnetic linear dichroism (MLD) - was observed in the ferromagnetic semiconductor (Ga,Mn)As. This discovery not only provided a new experimental tool for the investigation of in-plane magnetization dynamics in (Ga,Mn)As using light at normal incidence, but it also motivated the development of experimental techniques for the measurement of second-order MO effects in general. In this paper we compare four different experimental techniques that can be used to measure MLD and to separate it from experimental artifacts. We show that the most reliable results are obtained when the harmonic dependence of MLD on a mutual orientation of magnetization and light polarization plane is used together with the in-situ rotation of the sample followed by the magnetic field-induced rotation of magnetization. Using this technique we measure the MLD spectra of (Ga,Mn)As in a broad spectral range from 0.1 eV to 2.7 eV and we observe that MLD has a comparable magnitude as polar MOKE signals in this material.

## I. INTRODUCTION

Magneto-optical (MO) spectroscopy is a powerful tool for investigating basic properties of various magnetic and non-magnetic materials such as the electronic structure, magnetic anisotropy, spin population and magnetic excitations.[1-7] The first-order magneto-optical Kerr effect (MOKE) is usually used for this purpose due to the relatively large signals that it produces and its sensitivity to both in-plane (longitudinal and transverse Kerr effect) and out-of-plane (polar Kerr effect) positions of magnetization, respectively.[8-11] Although second-order (or quadratic) MO effects were investigated thoroughly by the magneto-optical community,[3,9,12-15] they were usually disregarded in materials research because they are observable only in materials with in-plane magnetization and usually lead to much smaller signals compared to the first-order MOKE.[3,9,13,14,16] In 2005 a giant quadratic MOKE effect - magnetic linear dichroism (MLD) - was observed in ferromagnetic (FM) semiconductor (Ga,Mn)As.[5] The discovery of a quadratic magneto-optical effect with a magnitude comparable to the first-order polar Kerr effect (PKE),[4,5,17] established MLD as a legitimate tool for investigating different material properties and interesting physical phenomena.[4,5] For example, for the normal incidence of light, PKE and MLD are sensitive to the out-of-plane and in-plane projections of the magnetization, respectively.[4] Consequently, the simultaneous measurement of the polarization rotation due to PKE and MLD enabled a reconstruction of the real space magnetization trajectory induced by the impact of a laser pump pulse in (Ga,Mn)As.[4,17,18] MLD can also be perceived as anisotropic magnetoresistance (AMR) at

---

[a)]Electronic mail: nada.tesarova@mff.cuni.cz.

finite frequencies, as both the DC and AC phenomena are even in magnetization, as was shown for photon energies near 1 eV.[19] This is analogous to PKE, which can be viewed as the AC anomalous Hall effect[20] that is odd in magnetization. Moreover, the sensitivity of MLD to the energy states that are responsible for magnetic order in FM semiconductors[5] makes MLD spectroscopy a very promising tool for providing new insights into these materials.

The reliable experimental measurement of the polarization change due to MLD is, however, a challenging task. While the experimental techniques for measurements of first-order MOKE are well-established and relatively easy to use, their direct application to MLD is not possible. For example, the magnitude of PKE is determined by measuring the light polarization rotation induced by the out-of plane oriented magnetization $M$. Since PKE is odd in magnetization (as it is proportional to $M$), the MO signals measured for parallel and antiparallel orientations of $M$ with respect to the direction of the incident beam, which are set by the direction of a saturating magnetic field, should have a same magnitude but opposite sign. All the possible non-magnetic artifacts, which are typically present in the measured signals, can thus be readily removed by determining the difference between the signals measured at $+M$ and $-M$. On the contrary, MLD is an even function of the magnetization (as it is proportional to $M^2$, [Ref. 12, Ref. 3]). This means that the 180° magnetization reorientation leads to the same MO signal[4] and, consequently, it is not an easy task to separate the real MO signal from polarization artifacts. In this paper we compare several experimental procedures that enable polarization artifact removal and show that the most reliable results are obtained by our novel experimental technique that employs the polarization dependence of MLD. Using this technique we measured the MLD spectra of the archetypical FM semiconductor (Ga,Mn)As in a broad spectral range (0.1 – 2.7 eV), which covers all the optical transitions from states that could be responsible for the FM order in this semiconductor.[2,5]

This paper is divided into six parts. In Sec. II, we first introduce the phenomenological description of MLD and derive its dependence on the polarization of the incident light. In Secs. III and IV we describe the different experimental techniques that we use to measure MLD in (Ga,Mn)As samples together with some interesting technical details of our apparatus. Finally, in Sec. V we present and discuss the achieved results. A detailed mathematical description of the methods is shown in the Appendix.

**II. PHENOMENOLOGICAL DESCRIPTION OF MLD**

In general, MLD is a second-order MO effect which is caused by a different (complex) index of refraction for light polarized parallel and perpendicular to magnetization orientation. MLD was originally observed in transmission, as a dichroism of linearly polarized light induced by the presence of a magnetic field or magnetization.[12] The difference in absorption for light polarized parallel and perpendicular to the magnetization orientation leads to a rotation of the polarization plane of linearly polarized light.[12] The same name was subsequently adopted also for the MO effect in the near-normal reflection geometry[4,5,15,21-23] where the rotation of linearly polarized light (or the change of its ellipticity) is caused by the different refraction indices for two orthogonal linear polarization components of light. We note that MLD is analogous to magnetic linear birefringence (MLB) - or Cotton-Mouton or Voigt effects, which are observed in the transmission geometry.[12,13,24]

In this article we will concentrate on the rotation of light polarization induced by MLD for light reflected at normal incidence from a sample with in-plane magnetization. The definition of the MLD signal in this context is the following:

$$MLD \ [\text{rad}] = \frac{1}{2} \frac{I_R^{\parallel} - I_R^{\perp}}{I_R^{\parallel} + I_R^{\perp}}, \qquad (1)$$

where $I_R^{//}$ and $I_R^{\perp}$ are the intensities of the reflected light polarized parallel and perpendicular to the magnetization, respectively. Equation (1) can be written equivalently in terms of the reflection coefficients $r_{//}$ and $r_{\perp}$:

$$MLD\ [rad] = \frac{1}{2}\frac{r_{\parallel}^2 - r_{\perp}^2}{r_{\parallel}^2 + r_{\perp}^2}. \tag{2}$$

The sign as well as the magnitude of the MLD signal are sensitive to the polarization orientation of the incident light. The polarization dependence of the MLD can be analytically calculated using the trigonometric relation between the magnetization orientation in the sample plane, given by the angle $\varphi_M$, and the incident and reflected light polarizations, given by the angles $\beta$ and $\beta'$ [see Fig. 1(a) for the angle definition]. The polarization rotation $\Delta\beta$ ($\Delta\beta \equiv \beta' - \beta$) due to MLD can be expressed as (see Ref. 4 for more details):

$$tan(\Delta\beta) = \frac{(r_{\parallel}-r_{\perp})tan(\varphi_M - \beta)}{r_{\parallel} + r_{\perp} tan^2(\varphi_M - \beta)}, \tag{3}$$

Assuming a small rotation of light polarization, i.e., $r_{//}/r_{\perp} \approx 1$, we obtain:

$$\Delta\beta = P^{MLD} sin[2(\varphi_M - \beta)], \tag{4}$$

where $P^{MLD} = 0.5(r_{//}/r_{\perp} - 1)$ is the MLD magneto-optical coefficient. Equation (4) shows that the MLD is zero when the incident polarization is parallel or perpendicular to magnetization orientation (i.e., $\varphi_M - \beta = 0°$ or $90°$, respectively). On the other hand, the rotation of light polarization is maximized when the angle between the magnetization and the incident polarization is $\pm 45°$. In this case, the magnitude of the polarization rotation is given solely by the MLD magneto-optical coefficient $P^{MLD}$ and Eqs. (4) and (2) are equivalent if we again assume $r_{//}/r_{\perp} \approx 1$:

$$P^{MLD} = \frac{1}{2}\left(\frac{r_{\parallel}}{r_{\perp}} - 1\right) = \frac{1}{2}\frac{r_{\parallel} - r_{\perp}}{r_{\perp}}\frac{(r_{\parallel}+r_{\perp})}{(r_{\parallel}+r_{\perp})} \approx \frac{1}{2}\frac{r_{\parallel}^2 - r_{\perp}^2}{r_{\parallel}^2 + r_{\perp}^2}. \tag{5}$$

### III. EXPERIMENTAL TECHNIQUES

We use several different experimental configurations to measure the MLD in (Ga,Mn)As. The near normal reflection geometry is employed in all configurations – the angle between the incident beam direction and the sample normal, $\psi$, did not exceed $6°$ (see Fig. 1).

The polarization rotation due to MLD can be measured directly from its definition using Eq. (1). As a first step, the magnetization is oriented by applying a saturating external magnetic field in the sample plane. The reflected light intensity is then measured for the incident light polarization parallel and perpendicular to the magnetization orientation, respectively, and the magnitude of MLD is computed from Eq. (1). Although this method works in principle, it is usually necessary to use other experimental techniques that enable more sensitive measurements of the small polarization rotations – especially in (Ga,Mn)As where the MLD magnitude typically does not exceed 1 mrad.

The most common method for determining the MLD-related MO signal is the measurement of the hysteresis loops for magnetic field sweeps in the sample plane.[3,5,9,14,15] In Fig. 1(b) we show the typical setup used for this kind of experiment. As mentioned in the introduction and as can be also seen from Eq. (4), the MLD signals are the same for two opposite orientations of magnetization, making the hysteresis loop measurement impossible for 180° reorientation of magnetization. However, MLD hysteresis can be measured in samples with fourfold magnetocrystaline anisotropy.[3,9,14-16] In (Ga,Mn)As the fourfold symmetry is a consequence of the competing uniaxial and cubic magnetic anisotropies, resulting in four equivalent magnetization easy axes (instead of two, as in conventional ferromagnets).[5,23,25] Magnetic field sweeps in the sample plane thus result in the M-shaped MO signal reflecting the magnetization jumps among these four easy axes.[5,23,25] In order to measure such hysteretic signals, the incident polarization of light is set by the polarizer ($P_1$) to an "appropriate" orientation (the best orientation choice will be discussed in detail below) and the magnetization induced polarization rotation is detected by a polarization-sensitive optical bridge, which consists of a half wave plate ($\lambda/2$), polarizing beam splitter ($P_2$) and two detectors ($D_1$ and $D_2$). We note that although this setup is very sensitive to small rotations of the light polarization plane, a quantitative determination of the MLD magnitude from such measurements is not straightforward and will be discussed together with the obtained results later in the text.

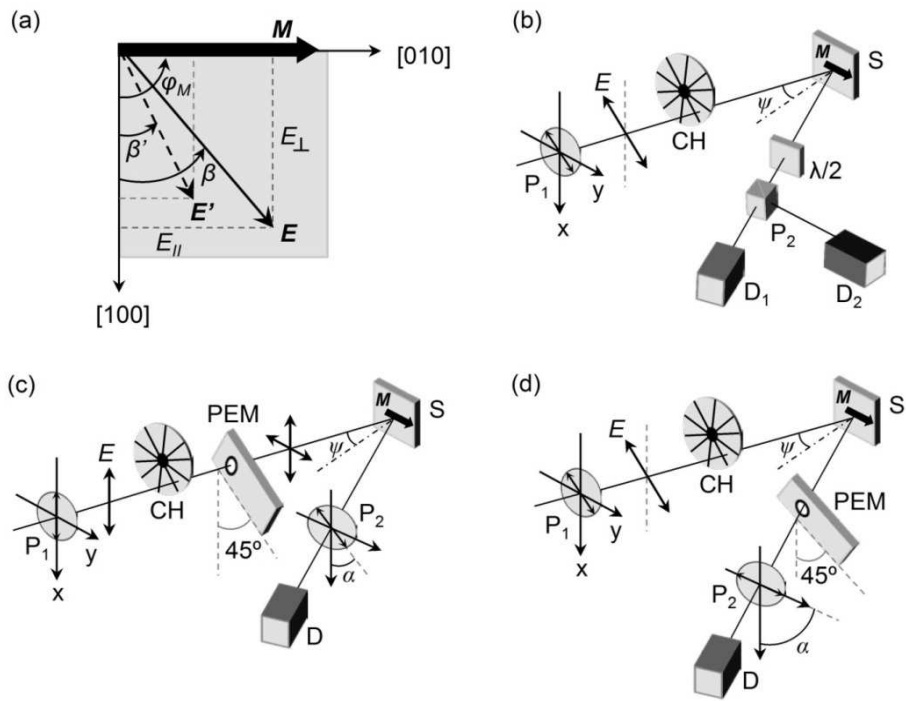

Fig. 1. Magnetization-induced rotation of light due to MLD and schematic illustrations of several experimental configurations used to detect MLD. (a) Different reflection coefficients for light polarized parallel ($E_\parallel$) and perpendicular ($E_\perp$) relative to magnetization orientation ($M$), at an angle $\varphi_M$, lead to a rotation of the polarization plane of linearly polarized light. The orientations of the incident $E$ and reflected $E'$ polarization planes are described by angles $\beta$ and $\beta'$, respectively. (b) The rotation of light polarization, which is set by polarizer ($P_1$), after reflection from sample (S) can be measured in a setup with a polarization-sensitive optical bridge that consists of half wave plate ($\lambda/2$), polarizing beam splitter ($P_2$) and two detectors ($D_1$, $D_2$). (c) and (d) Experimental configurations using a photoelastic modulator (PEM). During the measurement, polarizer ($P_2$) is oriented at $\alpha = 45°$ and $90°$ in part (c) and (d), respectively. The incident beam is modulated by mechanical chopper (CH), $\psi$ is the angle between the incident beam direction and the normal to the sample surface.

Another approach for MLD measurement is to use highly sensitive experimental techniques that modulate the light polarization.[12] Two experimental configurations using polarization modulation are shown in Fig. 1(c) and (d). This technique is based on modulating the relative phase of two orthogonal linear polarizations that pass through a photoelastic modulator (PEM). In Fig. 1(c), the polarization of incident light is set perpendicular (or parallel) to the fixed position of the magnetization. The optical axis of the PEM is oriented 45° with respect to the incident polarization, so it modulates the phase difference $\delta$ between the $x$ and $y$ components of the polarization periodically: $\delta = \delta_0 \sin(\omega_{PEM} t)$, where $\delta_0$ is the dynamic retardance amplitude and $\omega_{PEM}$ is the natural resonant frequency of the PEM.[26] When $\delta_0 = \pi$, the polarization of the light transmitted through the PEM is changed from vertical to horizontal at a frequency $2\omega_{PEM} = 2\pi \times 100$ kHz.[26] These two perpendicular polarizations are reflected from the sample with different (complex) amplitudes, producing different projections on the optical axis of the subsequent linear polarizer ($P_2$). The intensity of light $I_D$ at the detector [D in Fig. 1(d)] is derived in the Appendix and can be expressed as:

$$I_D = \frac{I_0}{2}(1 + \cos\delta \cos 2(\alpha + \Delta\beta) + \sin\delta \sin 2(\alpha + \Delta\beta)\sin\Delta\eta), \tag{6}$$

where $I_0$ is the intensity of the incident light, $\Delta\beta$ represents the polarization rotation due to MLD, $\alpha$ is the angle that describes the rotation of the $P_2$ optical axis from the $x$-axis, and $\Delta\eta = \eta_y - \eta_x$ is the magnetization induced phase shift of the two reflected orthogonal polarizations, which causes the ellipticity of light.[12] In order to detect and subsequently unambiguously separate the MLD induced rotation and ellipticity, we set $\alpha = 45°$. We can now rewrite Eq. (6) as a series of harmonic terms with Bessel function coefficients, focusing on the first four terms in the expansion to obtain:

$$I_D \approx \frac{I_0}{2}(1 - J_0(\delta_0)\sin(2\Delta\beta) + 2J_1(\delta_0)\cos(2\Delta\beta)\sin(\Delta\eta)\sin(\omega_{PEM} t) - 2J_2(\delta_0)\sin(2\Delta\beta)\cos(2\omega_{PEM} t)), \tag{7}$$

where $J_n(\delta)$ is an $n$-th order of the Bessel function. The overall intensity of the light is modulated by a mechanical chopper at approximately 1 kHz, and provides the average or dc[26] intensity of the radiation. The output of the detector consists of three frequency components that are processed by lock-in amplifiers; a $I(0)$-component which is detected at the chopper frequency, and the odd and even frequency components of $\omega_{PEM} - I(\omega_{PEM})$ and $I(2\omega_{PEM})$, respectively (see Appendix for complete analysis). We note that Eq. (7) is only approximate, as each of the components has its own sensitivity given by the detection-amplification system, caused mainly by the detector/amplifier rolloff,[26] which is calibrated for all our measurements. Although this equation clearly shows that in principle we are able to detect not only the rotation of light polarization (term containing $sin(2\Delta\beta)$ at frequency $2\omega_{PEM}$), but also its ellipticity (term containing $sin(\Delta\eta)$ at frequency $\omega_{PEM}$), we will concentrate only on the MLD induced rotation of light in the following analysis. Assuming small rotation angles (i.e., $sin 2(\Delta\beta) \approx 2\Delta\beta$ and $J_0(\delta_0)sin 2(\Delta\beta) \ll 1$), we can write the expression for the magnetization induced polarization rotation as:

$$\Delta\beta = -\frac{I(2\omega_{PEM})}{I(0)} \frac{1}{4C_2 J_2(\delta_0)}, \tag{8}$$

where $C_2$ is a constant given by the frequency-dependent sensitivity of the detection system. In order to obtain the magnitude of $\Delta\beta$, i.e., the MLD coefficient $P^{MLD}$, one needs to perform a calibration procedure that sets the value of $C_2 J_2(\delta_0)$ (see the Appendix for a detailed description of the calibration procedure).

A modification of the experimental setup shown in Fig. 1(c) is obtained when the PEM is placed after the sample, as in Fig. 1(d). In this case, light is polarized at 45° with respect to vertical after passing through polarizer ($P_1$) so that the angle between the magnetization and the incident polarization is 45°. Analogously to the previous case, the light acquires ellipticity and rotation after being reflected from the sample. The polarization state of light is subsequently analyzed by the PEM, which modulates the relative phase between the two orthogonal polarizations – the unaltered incident polarization, which is parallel to the PEM optical axis, and the magneto-optically induced polarization, which is perpendicular to the PEM optical axis. The linear polarizer (P$_2$), which is placed after the PEM, is oriented at $\alpha = 90°$ (i.e., 45° with respect to the PEM optical axis) and mixes the two orthogonal polarization components exiting the PEM. Applying a similar analysis as for the previous setup (see the Appendix), we obtain the same mathematical expression for MLD, given by Eq. (8). In order to calibrate the measured signal we perform an in-situ calibration technique developed in Ref. 1. Here the PEM and P$_2$ are rotated as a single unit by a known angle, producing a well-defined signal at the $2\omega_{PEM}$ frequency. This signal is then used to calibrate the polarimetry system.[1] It is worth noting that this experimental setup is extremely sensitive to the orientation of the reflected polarization with respect to the PEM, and that even miniscule changes in the alignment due to non-magnetic artifacts can lead to spurious signals. In order to circumvent such experimental artifacts, one needs to measure the *relative* change of the signal due to the sample's magnetization. This can be done either by hysteresis loop measurements, i.e., by changing the magnetization position in the sample plane by an application of the in-plane external magnetic field, or by a physical rotation of the sample, i.e., by changing the magnetization position without an external magnetic field. In this paper we describe a technique where both these approaches are combined to get the most reliable MLD-related signals. A detailed experimental procedure is discussed later in the text.

## IV. EXPERIMENTAL DETAILS

The experiments are performed on two 20 nm thick (Ga,Mn)As samples with a nominal Mn concentration of 3% and 7%. The samples are grown on the GaAs(001) substrates by low temperature molecular beam epitaxy. The growth conditions and post-growth annealing are optimized for both samples in order to get as close as possible to the intrinsic properties of idealized, uniform and uncompensated (Ga,Mn)As epilayers (see Ref. 6 for more information). The Curie temperature ($T_C$) of the 3% and 7% Mn samples are 77 K and 159 K, respectively. The magnetic anisotropy of the samples was studied by the superconductive quantum interference device (SQIUD), showing the four equivalent, non-perpendicular easy axis orientations of the magnetization in the sample plane.[27]

In order to probe these samples over a broad energy range, the MLD measurements are done in two collaborating laboratories at Charles University in Prague and University at Buffalo, with the former using visible photon energies above 1.2 eV and the latter concentrating on lower, infrared photon energies below 1.2 eV. The MLD is measured using discrete spectral lines from CO$_2$ (115 – 133 meV), CO (215 – 232 meV) and Ti-sapphire (1.63 eV) lasers and two distinct broadband light sources: a halogen lamp with a diffraction grating monochromator (Jobin Yvon Spex Model HR250) and the Xe lamp (Perkin-Elmer Cermax) with a double-pass CaF$_2$ prism monochromator (Perkin-Elmer Model 99).[28] The main advantage of the prism monochromator is that each wavelength is dispersed into a unique angle, which is not the case for the diffraction grating monochromator where a cut-off filter is used to remove the higher order diffraction peaks. Unlike typical arc lamps, which are housed in glass (limiting their usage to below 2.5 $\mu$m), our Xe lamp is equipped with a sapphire window that enables access to longer wavelengths. Depending on the light wavelength, different sets of optics are used. In the 10.6 – 2 $\mu$m wavelength range (115 – 620 meV), we use a ZnSe PEM (II/ZS50, Hinds Instruments) and BaF$_2$ holographic wire-grid polarizers. A fused silica PEM (I/FS50, Hinds Instruments) and calcite Glan-Taylor polarizers are used in the 2 – 0.46 $\mu$m wavelength range (620 – 2 700 meV). To avoid interference effects caused by multiple reflections within the PEM when coherent laser light is used, the ZnSe PEM is tilted forward 25° and the fused silica PEM crystal is wedged. Two different cryostats are used: a superconducting magneto-optical cryostat (Cryo Industries), reaching temperatures down to 6 K and magnetic fields up to 7 T and an optical cryostat (Janis Reasearch) reaching temperatures down to 8 K with a separate electromagnet (HV – 4H, Walker Scientific) producing a magnetic field up to 2 T. In order to rotate the sample at low temperatures ($T \sim 15$ K), a special copper sample holder is constructed for the 7 T magneto-optical cryostat, enabling a complete (360°) rotation of the sample around an axis parallel to the incident radiation direction. The sample rotation is achieved by two kevlar threads wrapped around the sample holder and around two brass cylinders that are placed at the top of the sample stick. The kevlar threads pass through separate vacuum feed-throughs inside the cryostat. The actual rotation of the sample holder is done by rotating the brass cylinders, which increases the tension on one thread while decreasing it on the other. A schematic illustration and the photograph of the sample holder are shown in Fig. 2.

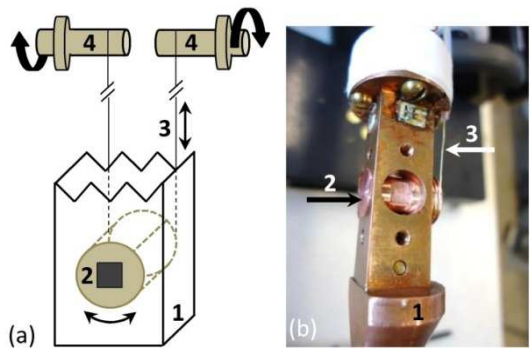

Fig. 2. (Color online) Rotating sample holder. (a) Schematic illustration of the coldfinger (1), from the front side, with the sample holder (2) and sample (black square). The arrows indicate how the sample holder is rotated when kevlar threads (3), which are wrapped around two brass cylinders (4), are pulled/loosened. (b) Photograph of the rotating sample holder. The sample is glued on the front side of the holder.

## V. RESULTS AND DISCUSSION

In Fig. 3 we show the measurement of MLD using the definition described by Eq. (1). In this experiment, the magnetization is oriented by a strong external magnetic field of 500 mT (that is well-above the saturation field) along the [010] crystallographic direction (see Fig. 1(a) for a definition of the coordinate system). The intensity of the reflected light with the polarization parallel ($I_R^{//}$) and perpendicular ($I_R^{\perp}$) to magnetization direction is measured [see Fig. 3(a)]. The most important aspect of this measurement is to keep the incoming intensity the same for both polarizations of light. In order to meet this condition we use a linear polarizer with a fixed orientation and a $\lambda/2$ Fresnel rhomb to rotate the incident light polarization. The $I_R^{//}$ and $I_R^{\perp}$ spectra, measured at 15 K, are shown in Fig. 3(a) for GaMnAs sample with 3% of Mn. The overall shape of the intensity spectra is dominated by the emission spectrum of the halogen lamp, and no apparent difference can be seen between the curves. The difference between $I_R^{//}$ and $I_R^{\perp}$ is more pronounced in Fig. 3(b), where we show the signal that is calculated from the measured data using Eq. (1). The same procedure is also performed at temperature $T = 150$ K, which is high above the sample Curie temperature, with no magnetic field applied. Despite the zero magnetic moment in the sample, our measurements show a nonzero difference signal also at $T = 150$ K which can be attributed to experimental artifacts in our system (e.g., a birefringence of the cryostat windows, a polarization-sensitive response of the photodetector, etc.). By subtracting the curves in Fig. 3(b) we obtain the MLD spectrum of the sample without the artifacts [see Fig. 3(c)]. The relatively large experimental error comes from the poor reproducibility of the MLD spectrum, caused mainly by the fact that a relatively small MLD signal is obtained by subtracting two large signals. In order to confirm the MLD-related origin of the measured spectrum, we probe the polarization dependence of the signal around its peak (~ 1.6 eV). The results that are shown in Fig. 3(d) are in a good agreement with the polarization dependence of MLD described by Eq. (4). It is worth noting that since there are two orthogonal polarizations used in order to measure the MLD signal, the incident light polarization $\beta = 45°$ corresponds to Eq. (1). The measured data are fitted by the Eq. (4), obtaining the MLD coefficient $P^{MLD} = 0.9987$.

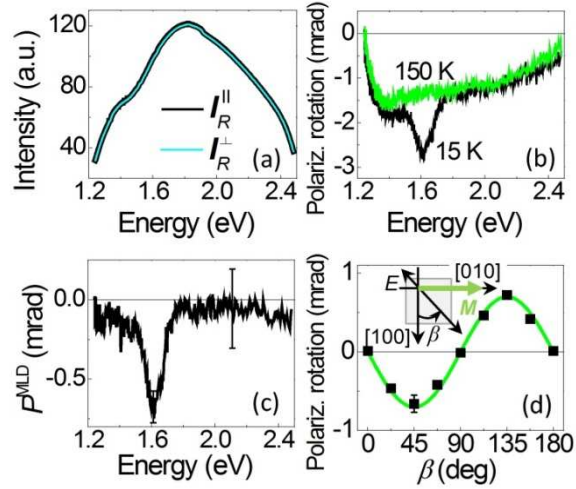

Fig. 3. (Color online) Measurement of MLD from the definition described by Eq. 1 in a $Ga_{1-x}Mn_xAs$ epilayer with nominal Mn doping $x = 3\%$ with the Curie temperature $T_C = 77$ K. (a) Spectral profile of reflected light intensity with polarization parallel ($I_R^{//}$) and perpendicular ($I_R^{\perp}$) to the magnetization at $T = 15$ K. (b) Polarization rotation determined from the data shown in (a) using Eq. (1) for $T = 15$ K. The same measurements were performed for a temperature above the Curie temperature with no external magnetic field. The data obtained for 150 K should not have any MLD features since $|M| = 0$, so any structure in these data are reproducible experimental artifacts. (c) MLD spectrum obtained by subtracting the curves shown in (b). (d) Polarization dependence of the MLD (black points) measured at photon energy 1.6 eV in the geometry shown in the inset, where the polarization rotation for $\beta = 45°$ corresponds to Eq. (1). The solid line is a fit by Eq. 4 with an MLD coefficient $P^{MLD} = 0.9987$ mrad. As expected, the polarization rotation signal is maximal at $\beta = 45°$, where $E_{\parallel} = E_{\perp}$ and zero when the incident polarization is perfectly perpendicular ($\beta = 0°$) or parallel ($\beta = 90°$) to $M$.

In Fig. 4(a) we show the hysteresis loops measured in the experimental setup depicted in Fig. 1(b) for the 3% Mn sample. A Ti-sapphire laser tuned to a photon energy of 1.62 eV is used to obtain a large MLD-related signal [see Fig. 3(c)]. The M-shaped hysteresis loops are a typical signature of four energetically equivalent magnetization easy axes in the sample plane [labeled "$M_1$" – "$M_4$" in Fig. 4(c)].[5,15,23,25] For a detailed understanding of the measured MO data, it is illustrative to perform the following analysis. Let us assume that by the application of a magnetic field the magnetization jumps from the easy axis (EA) $M_4$ to $M_1$. The orientation of magnetization in the sample plane is described by angles $\varphi_{M1} = \gamma - \xi/2$ and $\varphi_{M4} = \gamma + \xi/2$, respectively, where $\gamma$ is the position of the easy axes bisector and $\xi$ is their mutual angle [see Fig. 4(d) for a definition of the angles $\gamma$ and $\xi$]. According to Eq. (4), we can write the polarization rotation signals $\Delta\beta_1$ and $\Delta\beta_4$ for magnetization in $M_1$ and $M_4$, respectively, as:

$$\Delta\beta_1 = P^{MLD} sin2((\gamma - \frac{\xi}{2}) - \beta) \qquad (9)$$

$$\Delta\beta_4 = P^{MLD} sin2((\gamma + \frac{\xi}{2}) - \beta). \qquad (10)$$

The amplitude of the measured MO signal in the hysteresis loop is thus equal to $\Delta\beta = \Delta\beta_4 - \Delta\beta_1$:

$$\Delta\beta = 2P^{MLD}\cos 2(\gamma - \beta)\sin(\xi). \tag{11}$$

Equation (11) shows that the hysteresis loop amplitude is not only proportional to the MLD coefficient $P^{MLD}$ but also depends on the angle between two adjacent easy axes. The orientations of the EA are given by the overall magnetic anisotropy of the sample, which is quite complex in (Ga,Mn)As.[29,30] It consists of two competing contributions. The first one is the biaxial anisotropy along the [100] and [010] crystallographic directions, which originates from the cubic symmetry of the GaAs host lattice, and the second one is the uniaxial anisotropy along the [-110] crystallographic direction.[29,30] It is the uniaxial anisotropy that causes $\gamma = 135°$ in all (Ga,Mn)As samples and the maximum magnitude of the hysteresis loops is thus measured for $\beta = 45°$ or $135°$. We note that Eq. (11) shows the same periodicity as the "static" MLD-related signal [see Eq. (4)] - cf. Fig. 3(d) and Fig. 4(b).

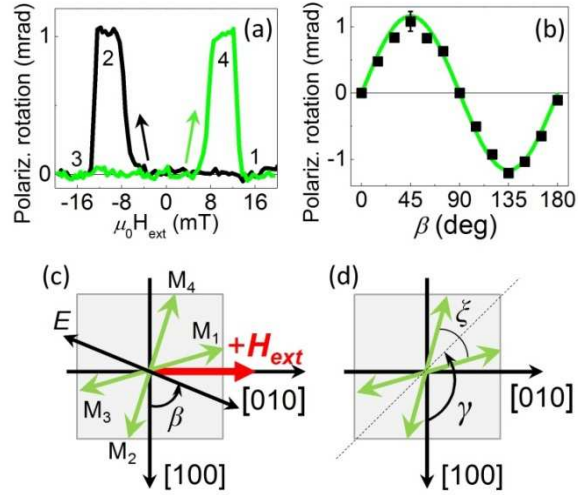

Fig. 4. (Color online) Rotation of light polarization measured at 15 K using the polarization-sensitive optical bridge. (a) *M*-shaped hysteresis loops are a signature of four energetically equivalent magnetization easy axes, which are schematically labeled $M_1$ - $M_4$ in (c). The number adjacent to the measured MO signal indicate the orientation of the magnetization along a particular easy axis. The photon energy is 1.62 eV and the incident orientation of linear light polarization $\beta = 45°$, where $\beta$ is depicted in (c). The measurement begins with a positive applied magnetic field causing the sample to be magnetized along $M_1$. As the external magnetic field is reduced and becomes negative (black curve), the magnetization jumps to $M_2$, producing a new rotation signal. When the applied magnetic field becomes more negative, the magnetization jumps into $M_3$, which leads to the same MO signal as in the case of magnetization in $M_1$. The green curve in (a) shows the polarization rotation signal when the applied magnetic field is swept from negative to positive values. Note that this techniques measures relative changes in the reflected light polarization as the magnetization switches from one EA to another; the absolute polarization rotation in (a) is arbitrary. (b) Hysteresis loop amplitude as a function of the light polarization orientation $\beta$ (black points) together with the fit by Eq. 11 (solid line) with parameters $P^{MLD}$ = 0.9987 mrad and $\xi = 60°$, where $\xi$ is the angle between adjacent easy axes, as shown in (d).

The major difference, however, is that the magnetization orientation ($\varphi_M$) and the corresponding MO coefficient ($P^{MLD}$) can be directly determined from the data shown in Fig. 3(d). On the contrary, the polarization dependence of the hysteresis loops [data in Fig. 4(b)] has to be supplemented by some additional independent measurement to obtain the information about the EA positions or $P^{MLD}$. For example, we can take the value $P^{MLD} = 0.9987$ from the fit of the data in Fig. 3(d) as the independent input and by fitting the data in Fig. 4(b) using Eq. (11) we can obtain the angle between the EA, $\xi = 60°$ and in turn the absolute orientation of EA in the sample plane: $\varphi_{M1} = 135° - \xi/2 = 105°$, $\varphi_{M4} = 135° + \xi/2 = 165°$, $\varphi_{M3} = \varphi_{M1} + 180°$, and $\varphi_{M2} = \varphi_{M4} + 180°$. We note that these results are in excellent agreement with the EA orientations obtained in this sample by independent time-resolved magneto-optical and SQUID measurements.[4,18,31,32]

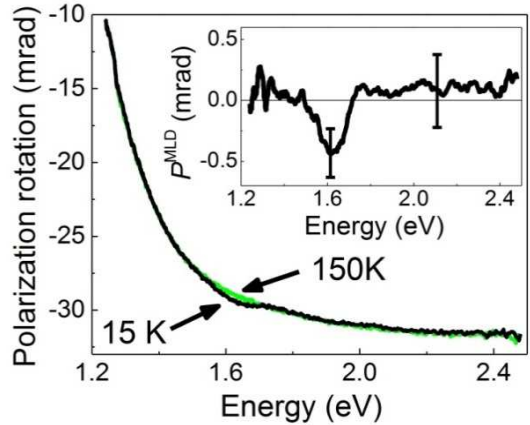

Fig. 5. (Color online) MLD measurement by the "temperature-corrected" PEM technique. The polarization rotation detected at temperatures below (15 K) and above (150 K) the sample Curie temperature are shown. Inset: Spectral dependence of MLD determined from the difference between the data depicted in the main panel.

In Fig. 5 we show the MLD measurement in the 3% Mn sample using the PEM experimental setup described in Fig. 1(c). We observe that the measured signal contains not only the MLD-related signal but also a strong background that is still present at temperatures above $T_C$. To remove it, we subtract the signals measured at 15 K and 150 K. The MLD spectrum obtained by this "temperature-corrected" PEM technique is shown in the inset of Fig. 5. The experimental error of this MLD signal is quite large due to the fact that we are subtracting two large signals with a magnitude of tens of miliradians in order to get a signal that is at least fifty times smaller. We note that, in principle, the background signal can be removed not only by heating up the ferromagnetic sample above its $T_C$ but also from a comparison of the signals measured at the same temperature in "similar" magnetic and non-magnetic samples (GaMnAs and GaAs in our case). We have verified that MLD spectra obtained by both approaches are similar, but the latter procedure provides less reproducible results. This is a consequence of small movements of the reflected beam within the experimental setup which cannot be avoided when the samples are interchanged. The "temperature-corrected" PEM technique could be also used to measure the MLD-related change of reflected light ellipticity [see Eq. (7)]. However, we observe that the detected signal at $\omega_{PEM}$, which is connected with the ellipticity, is not stable enough in time to enable a reliable comparison of the signals measured below and above $T_C$. This time instability of the signals, which is much more pronounced at $\omega_{PEM}$ (ellipticity) than at $2\omega_{PEM}$ (rotation), seems to be induced by slight variations of the PEM retardance, which is highly sensitive to temperature, due to drift in the laboratory ambient temperature.

Finally, we describe the measurement of MLD by our new experimental technique [Fig. 1(d)] where we combine the sensitivity of light polarization modulation using a PEM with the reduced background signal of hysteresis loop measurements [Fig. 1(b)]. The major advantage of the hysteresis measurements is that they reflect the *change* of the MO signal that is induced by the magnetization jump from one EA to another, and therefore "automatically" separate the measured signal from experimental artifacts. However, the common implementation of this technique can be applied only in samples with a sufficiently large angular separation $\xi$ between the adjacent magnetic easy axes – see inset in Fig. 4(c) and Eq. (11). We show below how this limitation can be eliminated if a rotation of the magnetization by an external magnetic field is supplemented by a rotation of the whole sample. Moreover, the analysis of the polarization state of the reflected light by the PEM can be performed in a much broader spectral range than in the case of the polarization-sensitive optical bridge employing a half wave plate. In the latter case, the wavelength range is extremely limited unless many different wave plates are used. Finally, both the magnetization-related change of light rotation and ellipticity can be measured simultaneously by detecting the signals at $2\omega_{PEM}$ and $\omega_{PEM}$, respectively. The limitation of this technique is that it can only be applied to samples with an in-plane magnetic anisotropy, where the positions of the magnetization EA are known from independent SQUID or pump-probe MO measurements.[4,32]

The measurement procedure itself consists of several steps which are schematically shown in Fig. 6(a). In the first step, we rotate the sample so that one of the EA is as close as possible to the direction of the external magnetic field ($H_{ext}$), which is horizontal in our case. By subsequent application of a strong magnetic field ($H_{ext} \sim 600$ mT) we "force" the magnetization to be aligned with this EA (as in case of four equivalent EA, the magnetization can be oriented in any of them). In the second step, we set $H_{ext}$ to zero and rotate the sample so that the magnetization is 45° away from horizontal. The position of the sample is now fixed and it is not changed during the actual measurement of the MLD. Next, we shine a light on the sample with a polarization plane along the magnetization orientation. Consequently, the reflected light should not experience any polarization rotation [see Eq. 4(a)] and, therefore, any signal measured at the reference frequency and its harmonics are just background artifacts that can be set to zero (e.g., by a small simultaneous rotation of the PEM and $P_2$). In the third step, we apply $H_{ext}$ that tilts the magnetization position until it is aligned with the direction of $H_{ext}$, which for the sample with 3% Mn is $\mu_0 H_{ext} > 20$ mT [see Fig. 6(b)]. Note that since $H_{ext}$ is large in this case, the orientation of M along $H_{ext}$ does not need to be along an EA. The 45° rotation of the magnetization within the sample plane causes one polarization component of the incident light to be aligned with M and the other, equal amplitude component to be perpendicular to M. As a result the rotation and ellipticity magnitudes change from minimal to a maximal values, with the changes being caused solely by MLD. We emphasize that the measured MO-signal is obtained without moving/rotating the sample, changing its temperature, or moving/rotating any optical elements. We find that unlike other changes, applying an external magnetic field is minimally disruptive to the reflected polarization. The 45° reorientation of the magnetization by the application of $H_{ext}$ yields a measured polarization rotation $\Delta\beta$ that is directly equal to $P^{MLD}$ [see Eq. (4a)]. We note that in a control experiment above the sample $T_C$ we did not observe any MLD-related MO signal [see Fig. 6(b)], in accordance with the negligibly small overall magnetic moment in the sample. The measured spectral dependence of $P^{MLD}$, where each data point is determined at a discrete photon energy using the procedure described above, is shown in Fig. 6(c). The error bars in the polarization rotation (~ 45 $\mu$rad) are mainly due to the uncertainty in the sample position, i.e., the EA position with respect to the magnetic field direction. The error bars in the ellipticity (~ 100 $\mu$rad) are larger due to more noise in the signal at $\omega_{PEM}$. It is worth noting, that some signal at $\omega_{PEM}$ is always present and cannot be zeroed as in the case of the rotation signal, and we refer to it as the background signal, which is not connected with the magnetization reorientation. The observed background ellipticity signal arises from optical components placed after the sample, e.g. cryostat windows, lenses, PEM etc. The ellipticity changes induced by the magnetization reorientation are typically on the order of the background signal noise, making the determination of MLD ellipticity less precise. In order to increase the sensitivity of the MLD measurements, one can rotate the sample 90° instead of 45° after the initial orientation of magnetization along an EA. In this case, the magnetization position is oriented 45° with respect to the incident polarization, causing a non-zero magneto-optical signal in rotation (and ellipticity) before the horizontal magnetic field is applied. This signal in rotation is zeroed deliberately by small rotation of the PEM and $P_2$ (as discussed above, some background ellipticity signal is always present and cannot be zeroed). The application of the external magnetic field will induce a 90° reorientation of the magnetization position, and now the polarization component of the incident light that was parallel with M becomes perpendicular to it and the other, equal amplitude component that was perpendicular to M becomes parallel to it. This results in a sign change of the MO signals, effectively *doubling* the measured step-like magneto-optical signal compared to the 45° rotation case, in accordance with Eq. (11).

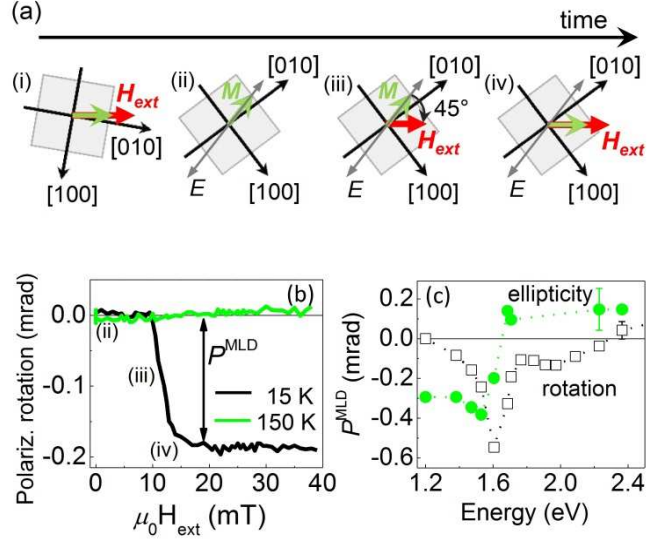

Fig. 6. (Color online) MLD measurement by the "rotation-corrected" PEM technique. (a) Schematic illustration of the individual steps performed in the measurement procedure. (i) Before the actual measurement of MLD, the sample easy axis (green arrow) is oriented along the saturating external magnetic field $H_{ext}$ (red arrow). (ii) In the next step, the magnetic field is turned off and the sample is rotated for 45° so that the magnetization M and the incident light polarization plane E are parallel, i.e., there is no polarization rotation due to MLD. (iii) and (iv) The position of the sample is fixed and the application of $H_{ext}$ leads to the magnetization reorientation and, consequently, to the light polarization rotation. (b) MLD signal produced by the change of the magnetization orientation relative to $E$, as indicated in (a), at temperatures below (15 K) and above (150 K) the sample Curie temperature, at a photon energy of 1.7 eV. (c) Spectral dependence of the light polarization rotation and ellipticity measured at $T = 15$ K.

We can now compare the MLD spectra measured in one sample using three different techniques: the determination of MLD using two separate measurements with orthogonal probing light polarizations [Fig. 3(c)]; "temperature-corrected" PEM technique (inset in Fig. 5); and the "rotation-corrected" PEM technique [Fig. 6(c)]. We see that the main characteristic feature of the MLD spectrum – the peak at ~ 1.62 eV – is present at the same position and with a similar magnitude in all the spectra. However, the major difference is in the spectral regions where only weak MLD-related signals are detected and, therefore, where in the first two methods a subtraction of two similar signals leads to a large uncertainty in the magnitude, and even in the sign, of the MLD signal [see the spectral ranges around 1.3 eV and 2.4 eV in Fig. 3(c) and in the inset in Fig. 5]. This uncertainty is, in principle, not present in the "rotation-corrected" PEM technique. To further illustrate the differences in the sensitivity of the experimental techniques, we show in Fig. 7 the MLD spectra measured in (Ga,Mn)As epilayers with 3% and 7% Mn content by the "temperature-corrected" (solid line) and the "rotation-corrected" (points) PEM techniques, respectively. We see that for the 7% Mn sample the subtraction of the measured data in the "temperature-corrected" technique appears to distort the measured spectral profile of MLD. We note that the magnitude of the MLD peaks in the visible and mid-infrared spectral regions is comparable with the magnitude of PKE measured in the identical samples.[4,17,18,31]

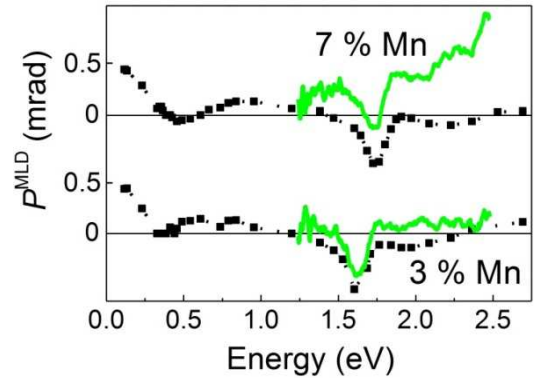

Fig.7. (Color online) Comparison of the spectral dependence of MLD measured in samples with 3% and 7% Mn concentration by two different experimental techniques. The solid line corresponds to MLD spectra measured by the "temperature-corrected" PEM technique, where rather strong experimental artifacts are present, and the points correspond to MLD spectra measured by the "rotation-corrected" PEM technique.

## VI. CONCLUSION

We present an overview of four different experimental techniques that can be used to measure the magnetic linear dichroism (MLD) in the ferromagnetic semiconductor (Ga,Mn)As. We show that the most reliable results are obtained using a new experimental technique that controls the angle between the sample magnetization and the polarization of probing light by rotating the sample and applying an external magnetic field. The main advantage of this technique is that it probes the magneto-optical signal that is directly connected with the ferromagnetic order in the sample while reducing artifacts from the experimental setup. In addition, this technique enables MLD measurements in samples with a strong uniaxial anisotropy, where only two EA, with 180° symmetry, are present. Using this technique we measure the MLD spectrum in a broad energy range from 0.1 eV to 2.7 eV in (Ga,Mn)As samples with 3% and 7% concentration of Mn atoms. We observe that the MLD is enhanced in the visible and the mid-infrared spectral regions, which is due to electronic interband transitions between the valence and conduction bands and intraband transitions within the valence band, respectively. The strong spectral features in the MLD spectrum might also bring light into the long lasting discussion about the character of the band structure in (Ga,Mn)As.

In conclusion, we would like to note that the experimental technique presented in this paper is not limited to ferromagnetic semiconductors such as (Ga,Mn)As, but it can be used to measure second-order magneto-optical effects in other ferromagnetic materials.

## VII. ACKNOWLEGEMENT

We wish to thank V. Novák, K. Olejník and M. Cukr for sample growth and characterization. We also thank T.J. Gruenauer and K.M. Cullinan for their work on designing and machining the rotating sample holder system that is used in the magneto-optical cryostat. We are grateful to K. Výborný for his helpful discussions about MLD from a theoretical point of view and A. Markelz for providing additional optical components.

This work was supported by the Grant Agency of the Czech Republic Grant No. 202/09/H041 and P204/12/0853 and the Grant Agency of the Charles University in Prague

Grant No. SVV-2012-265306 and No. 443011. We also acknowledge the financial support provided by the School of Physics, Faculty of Mathematics and Physics at the Charles University in Prague. Work done at University at Buffalo is supported by NSF-DMR1006078.

**APPENDIX**

In Fig. 1(c), the electric field is polarized along the *x*-axis when it is transmitted through the polarizer $P_1$. Taking advantage of the Jones matrix formalism, the electric field can be simply expressed in the linear basis as:

$$E_0 \begin{pmatrix} 1 \\ 0 \end{pmatrix}, \tag{A1}$$

where $E_0$ denotes the amplitude of the electric field. When passing through the PEM, with the optical axis tilted 45° from the *x*-axis towards the *y*-axis, we can write the electric field in the following form:

$$\tfrac{1}{2} E_0 \begin{pmatrix} 1 & -1 \\ 1 & 1 \end{pmatrix} \begin{pmatrix} 1 & 0 \\ 0 & e^{i\delta} \end{pmatrix} \begin{pmatrix} 1 & 1 \\ -1 & 1 \end{pmatrix} \begin{pmatrix} 1 \\ 0 \end{pmatrix}, \tag{A2}$$

where the periodic retardation of the PEM, $\delta$, is defined in the main text. The magnetized sample induces a rotation and a change in the ellipticity of the reflected light polarization, because of the different complex reflection coefficients for light polarized along the *x* and *y* axes. We can describe the magnetization-induced ellipticity by Jones matrix for a general phase retarder. The polarization rotation $\Delta\beta$ can be included in the projection of light polarization into the optical axis of the polarizer $P_2$, where the optical axis is oriented at the angle $\alpha$ with respect to the *x*-axis. The amplitude of the electric field vector ***E*** which is transmitted through $P_2$ is given:

$$\tfrac{1}{2} E_0 \begin{pmatrix} 1 + e^{i\delta} \\ 1 - e^{i\delta} \end{pmatrix} \begin{pmatrix} 1 & 0 \\ 0 & e^{i\Delta\eta} \end{pmatrix} \begin{pmatrix} \cos(\alpha + \Delta\beta) \\ \sin(\alpha + \Delta\beta) \end{pmatrix}, \tag{A3}$$

where $\Delta\eta = \eta_y - \eta_x$ is the polarization phase shift between the *x* and *y* polarization components producing the ellipticity. The overall intensity of light reaching the detector $I_D \sim |E|^2 = EE^*$ is given by Eq. (6) in the main text. The lock-in amplifiers demodulate the signal from the detector, which (in the first approximation) consists of three components:

$$I_D = q_0 I(0) + q_1 I(\omega_{PEM}) \sin(\omega_{PEM} t) + q_2 I(2\omega_{PEM}) \cos(2\omega_{PEM} t), \tag{A4}$$

where $I(0)$ is the intensity of the light modulated at the chopper frequency and $I(\omega_{PEM})$ and $I(2\omega_{PEM})$ are the light intensities modulated at $\omega_{PEM}$ and $2\omega_{PEM}$, respectively. $q_0$, $q_1$ and $q_2$ represent the sensitivities of the detection-amplification system for $I(0)$, $I(\omega_{PEM})$ and $I(2\omega_{PEM})$, respectively. Employing the Bessel functions, the detected signal components $I(0)$, $I(\omega_{PEM})$ and $I(2\omega_{PEM})$ can be written:

$$I(0) = \tfrac{I_0}{2}(1 + J_0(\delta_0)\cos 2(\Delta\beta + \alpha)), \tag{A5a}$$

$$I(\omega_{PEM}) = \tfrac{I_0}{2}(2J_1(\delta_0)\sin 2(\Delta\beta + \alpha)\sin(\Delta\eta)), \tag{A5b}$$

$$I(2\omega_{PEM}) = \tfrac{I_0}{2}(2J_2(\delta_0)\cos 2(\Delta\beta + \alpha)). \tag{A5c}$$

We note that the Eq. (7) in the main text was derived from Eq. (A4), (A5a), (A5b) and (A5c), respectively, assuming $\alpha = \pi/4$.

The polarization rotation $\Delta\beta$ can be determined as the ratio of $I(2\omega_{PEM})/I(0)$:

$$\frac{I(2\omega_{PEM})}{I(0)} = -C_2 \frac{2J_2(\delta_0)\cos 2(\Delta\beta+\alpha)}{1-J_0(\delta_0)\cos 2(\Delta\beta+\alpha)}, \tag{A6}$$

where the coefficient $C_2 = q_0/q_2$. In the small angle approximation and with the assumption that $\alpha = \pi/4$, Eq. (A6) leads to Eq. (8) in the main text.

The calibration of the experimental setup is needed in order to estimate the value of $C_2 J_2(\delta_0)$ in Eq. (8), thus obtaining $P^{MLD}$ from the measured rotation signal $\Delta\beta$. For this purpose, we replace the (Ga,Mn)As sample by a silver mirror and we first measure the signal for $\alpha = 0$ and subsequently for $\alpha = \pi/2$. The mirror does not induce any rotation of light polarization plane ($\Delta\beta = 0$), so the Eq. (A6) can be rewritten as:

$$\frac{I(2\omega_{PEM})}{I(0)} = \pm \frac{2C_2 J_2(\delta_0)}{1 \pm J_0(\delta_0)}, \tag{A7}$$

where the "+" sign corresponds to $\alpha = 0$ and "−" sign corresponds to $\alpha = \pi/2$. $J_0(\delta_0)$ can be determined by taking the ratio of $(I(2\omega_{PEM})/I(0))_{\alpha=0}$ and $(I(2\omega_{PEM})/I(0))_{\alpha=\pi/2}$. The value $C_2 J_2(\delta_0)$ can be obtained from Eq. (A7), using the calculated value of $J_0(\delta_0)$.

The similar mathematical analysis can be applied in case of PEM placed after the sample as shown in Fig. 1(d). In this case, the incident polarization of light is at 45° with respect to $x$:

$$\frac{E_0}{\sqrt{2}} \begin{pmatrix} 1 \\ 1 \end{pmatrix}. \tag{A8}$$

After the reflection from the magnetized sample, the electric field vector can be written in the form:

$$\frac{E_0}{\sqrt{2}} \begin{pmatrix} \cos(\frac{\pi}{4} + \Delta\beta) \\ e^{i\Delta\eta}\sin(\frac{\pi}{4} + \Delta\beta) \end{pmatrix}. \tag{A9}$$

The PEM modulates the phase difference between the components parallel and perpendicular to its optical axis and $P_2$, oriented at the angle $\alpha = 90°$, mixes them subsequently. The magnitude of the electric field vector after $P_2$ is given by a multiplication of Jones matrices of all the optical components:

$$\frac{E_0}{\sqrt{2}} \begin{pmatrix} \cos(\frac{\pi}{4} + \Delta\beta) \\ e^{i\Delta\eta}\sin(\frac{\pi}{4} + \Delta\beta) \end{pmatrix} \begin{pmatrix} 1+e^{i\delta} & 1-e^{i\delta} \\ 1-e^{i\delta} & 1+e^{i\delta} \end{pmatrix} \begin{pmatrix} 0 \\ 1 \end{pmatrix}. \tag{A10}$$

The overall intensity calculations lead to the same expression for MLD as in Eq. (8).